\documentclass[useAMS,usenatbib]{mn2e}
\usepackage{amssymb}
\usepackage{graphicx}
\newcommand{\dist}{1.15$''$}
\newcommand{\disterr}{1.15$\pm$0.27$''$}
\newcommand{\disterrnowav}{1.15$\pm$0.08$''$}
\newcommand{\wav}{\textsc{wavdetect}}
\title[Progenitor detection of SN 2007on]{On the detection of the progenitor of the type Ia supernova 2007on}

\author[Roelofs, Bassa, Voss \& Nelemans]{
Gijs Roelofs,$^{1}$\thanks{E-mail: groelofs@cfa.harvard.edu}
  Cees Bassa,$^{2}$ Rasmus Voss,$^{3,4}$ and Gijs Nelemans$^{5}$ 
  \\
  $^{1}$Harvard--Smithsonian Center for Astrophysics, 60 Garden Street,
  Cambridge, MA 02138, USA\\
  $^{2}$Department of Physics, McGill University, 3600 University Street, Montreal, QC H3A 2T8, Canada\\
  $^{3}$Max Planck Institute for Extraterrestrial Physics, Giessenbachstra\ss{}e, 85748, Garching, Germany\\
  $^{4}$Excellence Cluster `Universe', Technische Universit\"at M\"unchen, Boltzmannstra\ss{}e
2, 85748, Garching, Germany\\
  $^{5}$Department of Astrophysics, IMAPP, Radboud University
  Nijmegen, PO Box 9010, 6500 GL Nijmegen, The Netherlands\\
}

\begin{document}

\date{Accepted ... Received \today}

\pagerange{\pageref{firstpage}--\pageref{lastpage}} \pubyear{2008}

\maketitle

\label{firstpage}

\begin{abstract}
We present new \emph{Chandra} X-ray observations and detailed astrometry of the field of
the type Ia supernova 2007on, for which the detection of a likely progenitor in archival \emph{Chandra} data was recently reported.

No source is detected in the new \emph{Chandra} images, taken six weeks after optical maximum. We calculate a 90--99\% probability that any X-ray source near the position of the supernova (SN) is fainter than in the pre-outburst images, depending on the choice of aperture, which supports the identification of the archival X-ray source with the SN.

Detailed astrometry of the X-ray and new optical images, however, gives an offset between the supernova and the measured X-ray source position of \disterr. Extensive simulations show that the probability of finding an offset of this magnitude is
$\sim$1\%, equal to the (trial-corrected) probability of a chance alignment with any X-ray source in the field. This casts doubt on the identification of the X-ray source with the progenitor, although the scenario in which at least some of the observed X-rays are connected to the supernova may be the \emph{least unlikely} based on all available data.

After a brief review of the auxiliary evidence, we conclude that only future X-ray observations can shed further light on the proposed connection between the X-ray source and the progenitor of SN 2007on, and thus whether an accreting white dwarf scenario is truly favoured for this SN Ia.

\end{abstract}

\begin{keywords}
Supernovae -- binaries: close -- white dwarfs -- X-ray: binaries
\end{keywords}

\section{Introduction}
\label{introduction}

Type Ia supernovae are thought to be the result of thermonuclear explosion of carbon-oxygen white dwarfs as they reach or exceed the Chandrasekhar mass limit (see, e.g., \citealt{hillebrandt, leibundgut} for reviews). Broadly speaking, there are two main classes of models: the accreting models, in which the white dwarf more or less steadily accretes matter from a (hydrogen-rich) companion star \citep{whelan,nomoto}, and the merging models, in which two white dwarfs coalesce under the influence of angular momentum losses due to the emission of gravitational waves \citep{tutukov,webbink,iben}.

Deciding which of the scenarios contribute to the observed SN Ia rates is difficult from a purely theoretical point of view. One may, however, expect observable differences between the accreting and merging scenarios. In an effort to constrain the progenitor scenario for individual SNe Ia, \citet{vossnature} recently started a search for progenitor detections of newly reported, nearby SNe Ia in archival \emph{Chandra X-ray Observatory} images, arguing that the X-ray luminosities of progenitors in the accreting models are expected to be much higher than those of the progenitors in the merging models.

Their search was successful: the nearby type Ia supernova 2007on in NGC\,1404 \citep{pollas,morrell,galyam,immler} turned out to have an X-ray counterpart in a combined 75\,ks \emph{Chandra}/ACIS exposure, taken 4.5 years before the explosion. The X-ray counterpart was detected at the 5-$\sigma$ level, and within $0.9\pm1.3''$ of the optical supernova. Based on the low chance alignment probability of $\sim$1\%, it was claimed to be the likely progenitor, favouring an accreting white dwarf scenario over a merger scenario for 2007on \citep{vossnature}.

Given the importance of the detection of the progenitor to a type-Ia supernova, we obtained follow-up \emph{Chandra} observations (section \ref{ddtsection}) and performed accurate astrometry of the field of 2007on using existing and new imaging (section \ref{astrometrysection}). We briefly review the further evidence as to the nature of the X-ray source in section \ref{discussionsection}, and conclude with a summary of our results in section \ref{conclusionsection}.

\section{New results}

\subsection{\emph{Chandra} DDT observations}
\label{ddtsection}

\begin{figure*}
\includegraphics[width=\columnwidth]{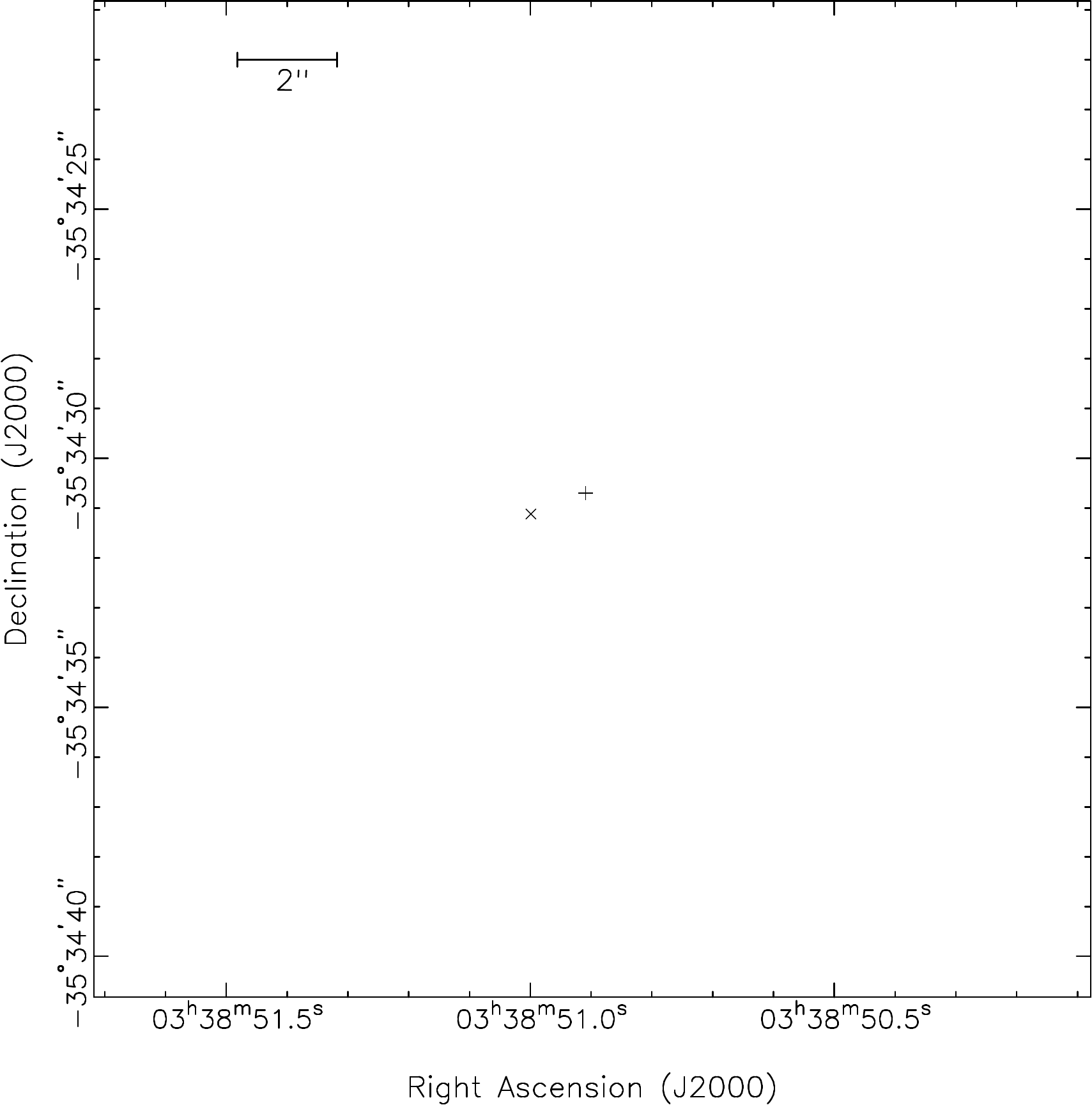}\hfill%
\includegraphics[width=\columnwidth]{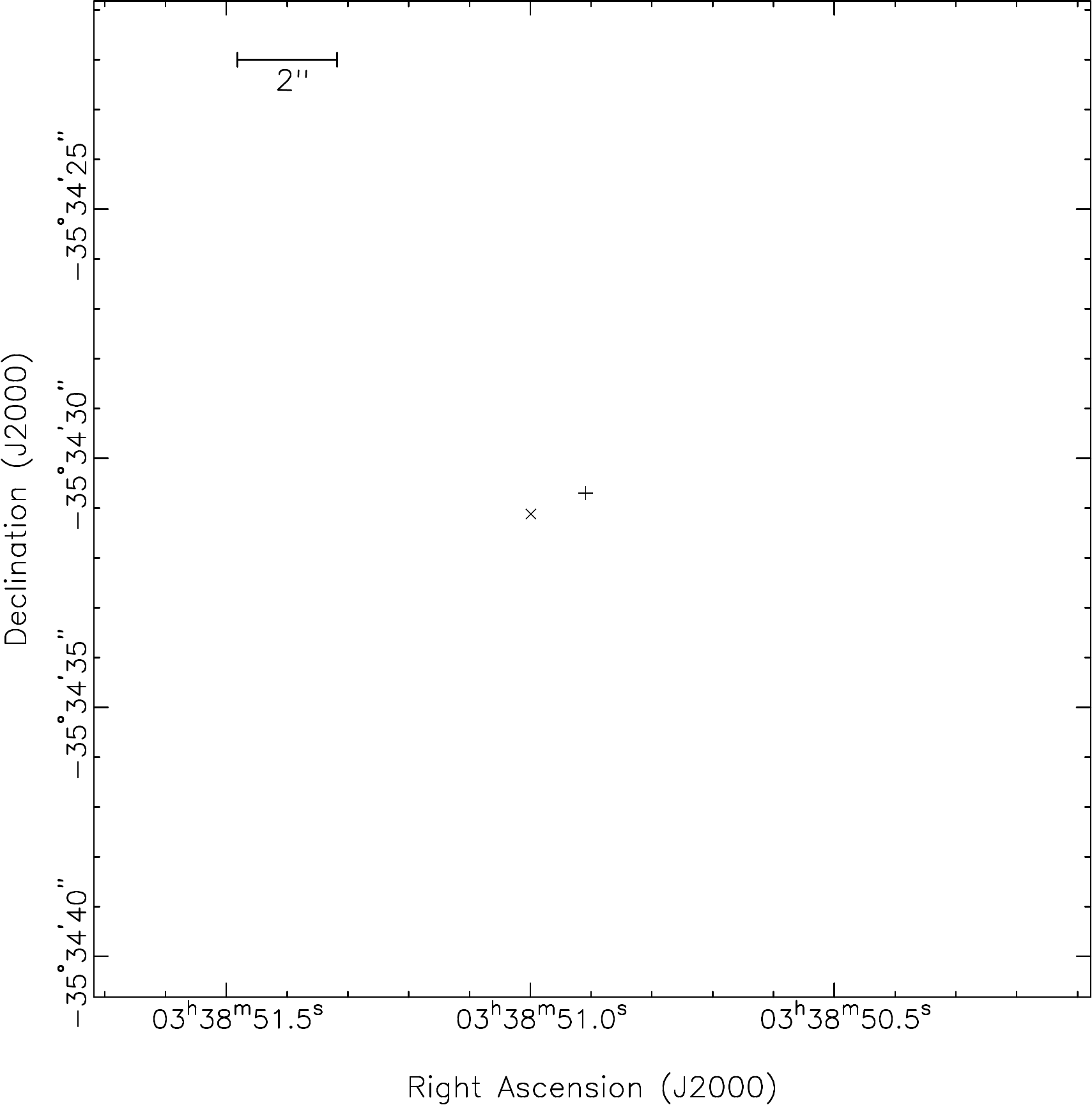}
\caption{Pre-outburst (left) and post-outburst (right) \emph{Chandra}/ACIS images of the region around 2007on. The gray-scale encoding of the squares indicates the integer number of counts in the corresponding pixels, white being zero and the darkest gray corresponding to 3 counts. The positions of the optical SN ($\times$) and the X-ray source as reconstructed by \wav\ ($+$) are indicated. Note that the effective area times exposure time of the post-outburst image is approximately 65\% of that of the pre-outburst image (see also Fig.~\ref{fig:probs}). The PSF in the post-outburst image should be slightly sharper since the source was aligned with the optical axis of \emph{Chandra}. The pixel randomizations applied by the \emph{Chandra} data processing pipeline were removed to produce these images.}
\label{fig:positions_chandra}
\end{figure*}

Following the detection of the likely progenitor to 2007on by \citet{vossnature}, we were granted 40\,ks of \emph{Chandra}/ACIS observations through the Director's Discretionary Time (DDT) programme. Due to pointing restrictions, 2007on was observed for 18.5\,ks on December 24, and 21.5\,ks on December 27, 2007. We used the ACIS-S instrument with the source positioned at the aimpoint on CCD S3. Being one of the two back-illuminated CCDs, S3 has the best sensitivity to soft X-rays; the archival X-ray source close to 2007on had been observed to be quite soft. Figure \ref{fig:positions_chandra} shows the pre-outburst image together with the new DDT observations.

The source appears to have become fainter, with 2 photons in a 1.0$''$ aperture that should contain $\sim$90\% of the flux, centered on the position of the X-ray source in the archival images, and counting only photons in the $0.3-8.0$\,keV energy range (for rough comparison, there were 17 such photons in a 1.0$''$ aperture in the progenitor image). Given the measured average background rate of 1.3 photons per 1.0$''$ aperture, the observed 2 photons suggest that the source may have disappeared altogether. A larger, 2$''$ aperture contains 10 counts (versus 22 in the pre-outburst image); this has a 4\% probability of happening by chance for a purely Poissonian background, or a measured 8\% probability of happening in a random aperture between 4$''$ and 20$''$ from the supernova in the post-outburst image. There is thus no convincing evidence that there is still a source in the follow-up observations. As expected the \wav\ source-finding algorithm from the standard \textsc{ciao} data-reduction package (version 4.0) does not find a source when run with the standard false-alarm probabilities of $1\times 10^{-6}$ per pixel.

\begin{figure}
\includegraphics[width=\columnwidth]{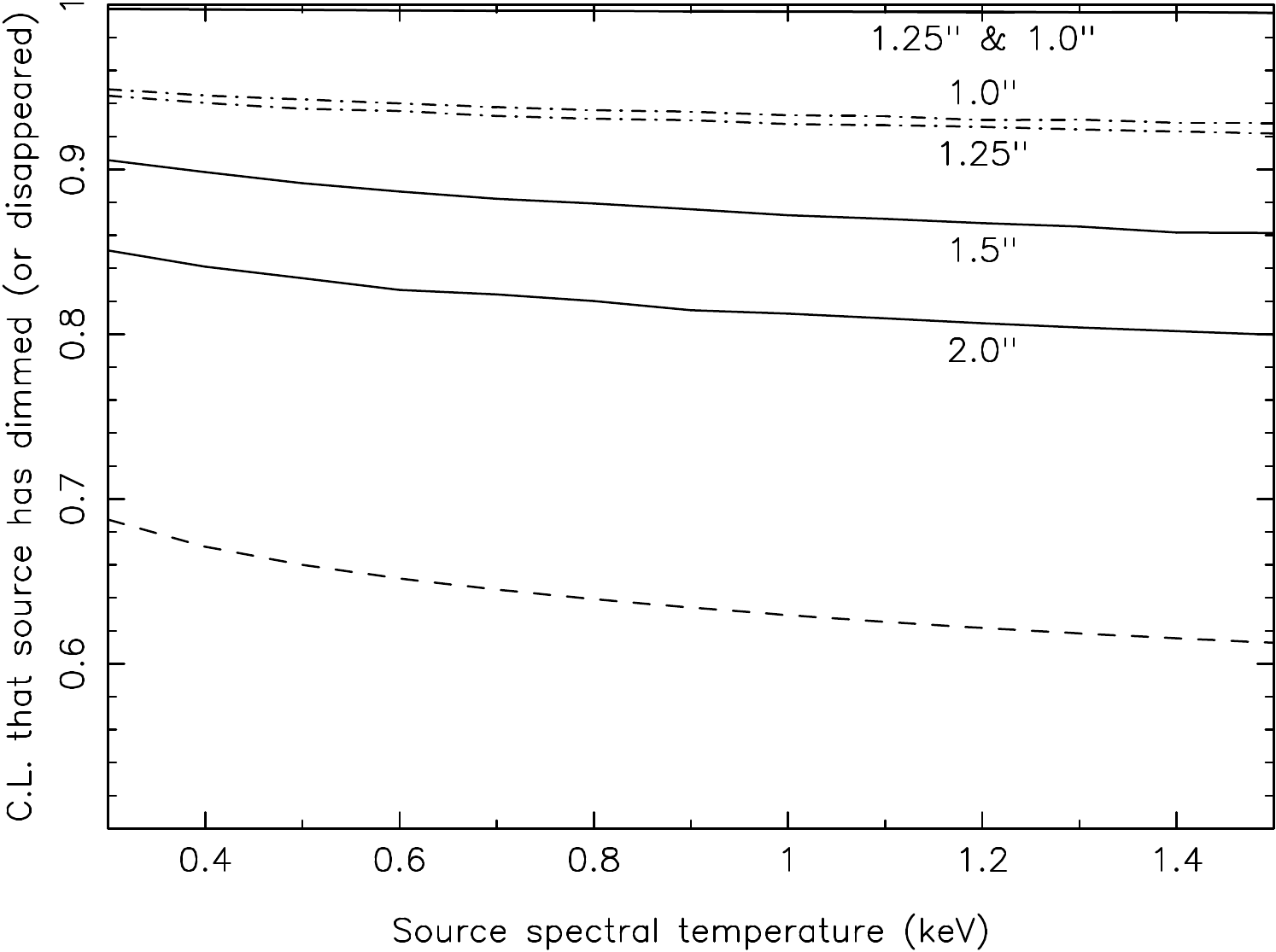}
\caption{Relative effective area times exposure time for our \emph{Chandra} DDT observations as a fraction of the pre-outburst value (dashed line), and the resulting confidence levels of the post-outburst source being fainter than the pre-outburst source (solid line, for different apertures centered on the pre-outburst X-ray source), as a function of the assumed temperature of the source. Dot-dashed lines indicate the confidence levels of the post-outburst source being fainter \emph{by at least a factor of 2}.}
\label{fig:probs}
\end{figure}

In order to be able to compare both epochs, we determined the relative effective collecting area times exposure time of the individual observations. Since part of the progenitor data were taken with one of the front-illuminated ACIS-I CCDs, which have different sensitivities especially in soft X-rays, and since furthermore the soft-X-ray sensitivity of ACIS has overall degraded significantly since Spring 2003 when the progenitor data were taken, the relative effective areas of the pre- and post-outburst data were expected to be a function of the assumed source spectrum. We thus worked out the relative effective areas times exposure times for the 2003 and 2007 epochs as a function of source temperature, which are shown in Fig.\ \ref{fig:probs}. As it turned out, the relative sensitivities were very much independent of the assumed source spectrum: the degradation in soft X-ray sensitivity of ACIS conspired with the lower soft X-ray sensitivity of the ACIS-I CCD used for part of the progenitor observations to produce an almost identical sensitivity-versus-energy curve for both epochs. This made comparison of both epochs considerably easier, since we did not have to worry about the source spectrum, or possible spectral changes between both epochs.

We performed a fully Poissonian Monte Carlo simulation to determine the probability of observing, by chance, the observed lower number of photons in the post-outburst image compared to the pre-outburst image, with the same (though unknown) underlying source luminosity. We did so for a range in apertures and, for completeness, also as a function of source spectral temperature, even though we had already concluded that the relative sensitivities of the observations in both epochs are quite insensitive to the assumed source temperature.

The results are also shown in Fig. \ref{fig:probs}. The confidence level for the X-ray source having dimmed increases with smaller aperture, to $>$99\% for apertures of radii $\leq$1.25$''$ centered on the position of the pre-outburst X-ray source. SN 2007on fell a distance 2.5$'$ and 1.7$'$ away from the optical axis in the pre-outburst ACIS-I and ACIS-S observations, respectively, so that a 1.0$''$ aperture should still contain approximately 80\% of the photons. However if one allows the source to be slightly extended or blended, a larger aperture may be appropriate, with a correspondingly lower confidence level. The slightly smaller PSF of the on-axis, post-outburst observations will make us underestimate the confidence level of the source having dimmed or disappeared.

We should mention that these results change slightly if we do the data analysis with the standard photon position randomization applied during the data reduction.\footnote{See \texttt{http://cxc.harvard.edu/ciao/threads/acispixrand/}} While there is no scientific justification for this randomization, it does shift several photons in the post-outburst image by 1 pixel, such that 4 instead of 2 fall within a 1.0$''$ aperture. This leads to confidence levels of the source having dimmed of 97\% and 91\% for 1.0$''$ and 1.25$''$ apertures respectively, while the confidence levels for the larger apertures go up slightly compared to the non-randomized results.

Finally, to test for variability of the source in the pre-outburst observations, insofar as the limited number of photons allows, we performed a Kolmogorov--Smirnov test against a uniform distribution (more accurately a set of uniform distributions with Poisson-distributed average rate) for both the ACIS-S and ACIS-I observations. The photon arrival times, and also those of subsets of only the soft or hard X-rays, were found to be perfectly compatible with uniform distributions, within the 1-sigma confidence intervals. There is thus no evidence for variability in the pre-outburst images.

\subsection{Astrometry}
\label{astrometrysection}

\subsubsection{Method}

\citet{vossnature} found an offset between the optical SN and the X-ray source of $0.9\pm1.3''$, entirely compatible with the X-ray source being the progenitor of the supernova. In order to verify that there is no significant offset between the two, we tried to improve on their astrometry.

We retrieved archival observations of NGC\,1404 
obtained with the Wide Field Imager (WFI) at the ESO 2.2-m telescope,
and of SN\,2007on taken November 30, 2007 with the ESO Faint Object Spectrograph and Camera (EFOSC) at the 3.6-m
telescope, both located at La Silla Observatory.

A 5-min $R$-band WFI image of NGC\,1404 was used to
astrometrically calibrate the EFOSC observation of the supernova. A
total of 76 stars from the two-micron all-sky survey (2MASS) coincided 
with the $8\arcmin\times16\arcmin$
field-of-view of a single WFI chip; 38 of these were not saturated
and appeared stellar. After iteratively removing outliers we obtained
an astrometric solution using 31 2MASS stars, yielding root-mean-square (rms)
residuals of $0\farcs091$ in both right ascension and declination.

A list of secondary astrometric standard stars was compiled by
measuring the positions of stars on the WFI image. These calibrated
positions were used to transfer the astrometry onto the 20\,sec
$R$-band EFOSC image of the supernova. The final astrometric
calibration used 71 stars common to both the WFI and the EFOSC image,
with rms residuals of $0\farcs092$ in right ascension and $0\farcs077$
in declination.

Based on this astrometric calibration, we determined the position of
the supernova in the EFOSC image to be
$\alpha_\mathrm{J2000}=03^\mathrm{h}38^\mathrm{m}50\fs999$ and
$\delta_\mathrm{J2000}=-35\degr34\arcmin31\farcs12$. The uncertainty
of the position on the EFOSC image is about $0\farcs006$ in both
coordinates, while the uncertainty on the absolute position is
$0\farcs13$ in right ascension and $0\farcs12$ in declination
(i.e.\ the quadratic sum of the uncertainties in the tie between the
2MASS catalog and the WFI image, the tie between the WFI and the
EFOSC images and the uncertainty of the supernova position in
the EFOSC image).

\begin{figure}
\includegraphics[width=\columnwidth]{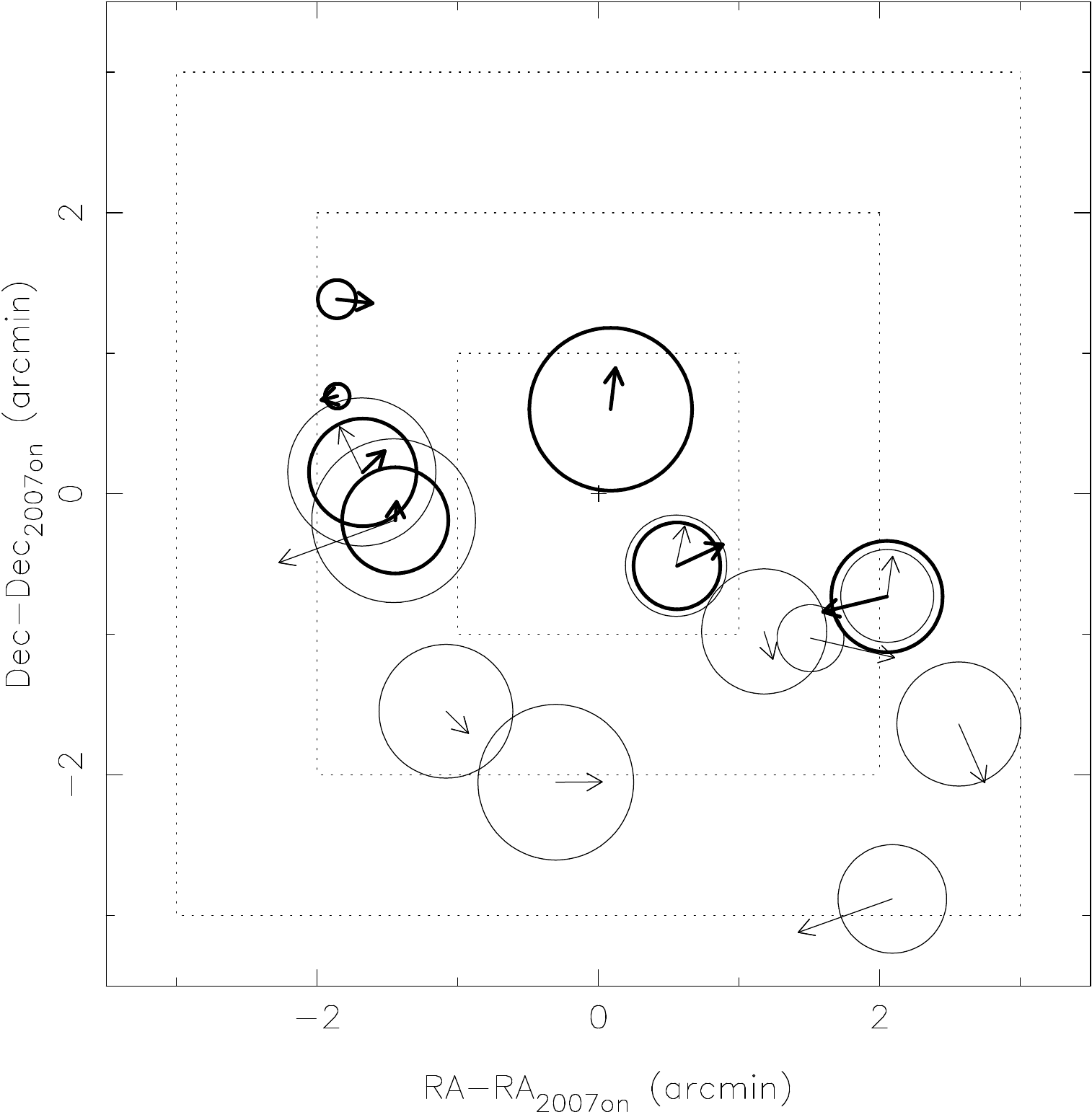}
\caption{Spatial distribution of the sources used for the X-ray to optical referencing. Right ascension and declination are relative to SN 2007on, marked with `+'. The \wav\ standard errors in the source positions are indicated by circles, while arrows indicate the displacement between the X-ray sources and their likely optical counterparts in the boresight-corrected image. Circle radii and arrow lengths are exaggerated by a factor 120, so that a 0.5$'$ arrow indicates a 0.25$''$ displacement. Circles and arrows drawn with thick and thin strokes show the reference sources for the ACIS-I and ACIS-S observations, respectively.}
\label{fig:offsets}
\end{figure}

In order to compare the optical position of the supernova with the
X-ray position of the candidate counterpart in the pre-outburst \emph{Chandra}
ACIS-I (ObsID 4174) and ACIS-S (ObsID 2942) observations, we determined
a boresight correction using X-ray sources (from \wav) which, by eye, had likely
optical counterparts in the EFOSC images. Objects that appeared elongated or blended in either the optical of X-ray image were rejected, and one rejection iteration of optical sources outside the 95\% confidence circles as given by \wav\ was applied. For the ACIS-I observation we thus
found 7 suitable X-ray sources, which
provided an offset to the raw X-ray positions of
$+0\farcs031\pm0\farcs026$ in right ascension and
$-0\farcs018\pm0\farcs025$ in declination. For the ACIS-S observation,
we found 10 X-ray sources, providing an offset of
$-0\farcs067\pm0\farcs049$ in right ascension and
$-0\farcs059\pm0\farcs048$ in declination. The boresight corrections thus appear to be very small. Fig.\ \ref{fig:offsets} shows the spatial distribution of the X-ray sources used for referencing the X-ray to the optical image.

Applying these corrections to the pre-outburst X-ray observations gives $\alpha_\mathrm{J2000}=03^\mathrm{h}38^\mathrm{m}50\fs909$ and
$\delta_\mathrm{J2000}=-35\degr34\arcmin30\farcs70$ for the absolute position of the X-ray source in the combined image, with uncertainties of $0\farcs25$ in right ascension and $0\farcs20$ in declination. For the relative positions of the optical SN and the X-ray source, the uncertainty in the referencing to the absolute ICRS frame is irrelevant, so the error in their relative positions is smaller. We find that the X-ray source is offset by
\disterr\ from the optical supernova, or \disterrnowav\ \emph{excluding} the error on the position of the X-ray source as given by \wav\ (i.e., only including the positional uncertainty of the optical SN relative to its reference stars, and the tie between the optical and X-ray reference stars), where we have conservatively added the errors on the boresight corrections in quadrature for want of a better method. The uncertainty in the position of the X-ray source, relative to its reference sources, is thus by far the dominant uncertainty, as expected.

\subsubsection{Implications}

To estimate the probability of such a misalignment happening by chance, i.e. with the X-ray source near 2007on still corresponding to the optical supernova, we performed a detailed Monte Carlo simulation in which we added artificial sources to the combined progenitor image from the \emph{Chandra} archive. We began by running \wav\ on the image to detect the sources in the image. From the source and smoothed background images returned by \wav, we filtered all pixels that had an average background within 10\% of the value measured around the progenitor (6.9 counts per 2$''$ aperture), and that had no detected source flux anywhere within five times this radius. We randomly picked coordinates from the qualifying pixels, took a random sub-pixel sampling, and added photons according to a bivariate normal distribution, until the number of photons in a 2$''$ aperture equalled 21, the number found in the pre-outburst images \citep{vossnature}. Since the PSF width in \emph{Chandra} images varies quite a bit with energy, we probably gained little by modelling the PSF with a more complicated function. Instead we ran the simulation for three reasonable widths of the distribution, choosing 0.8$''$, 1.0$''$ and 1.2$''$ radii of 86\% encircled energy, and doing 40,000 trials for each.

\begin{figure}
\includegraphics[width=\columnwidth]{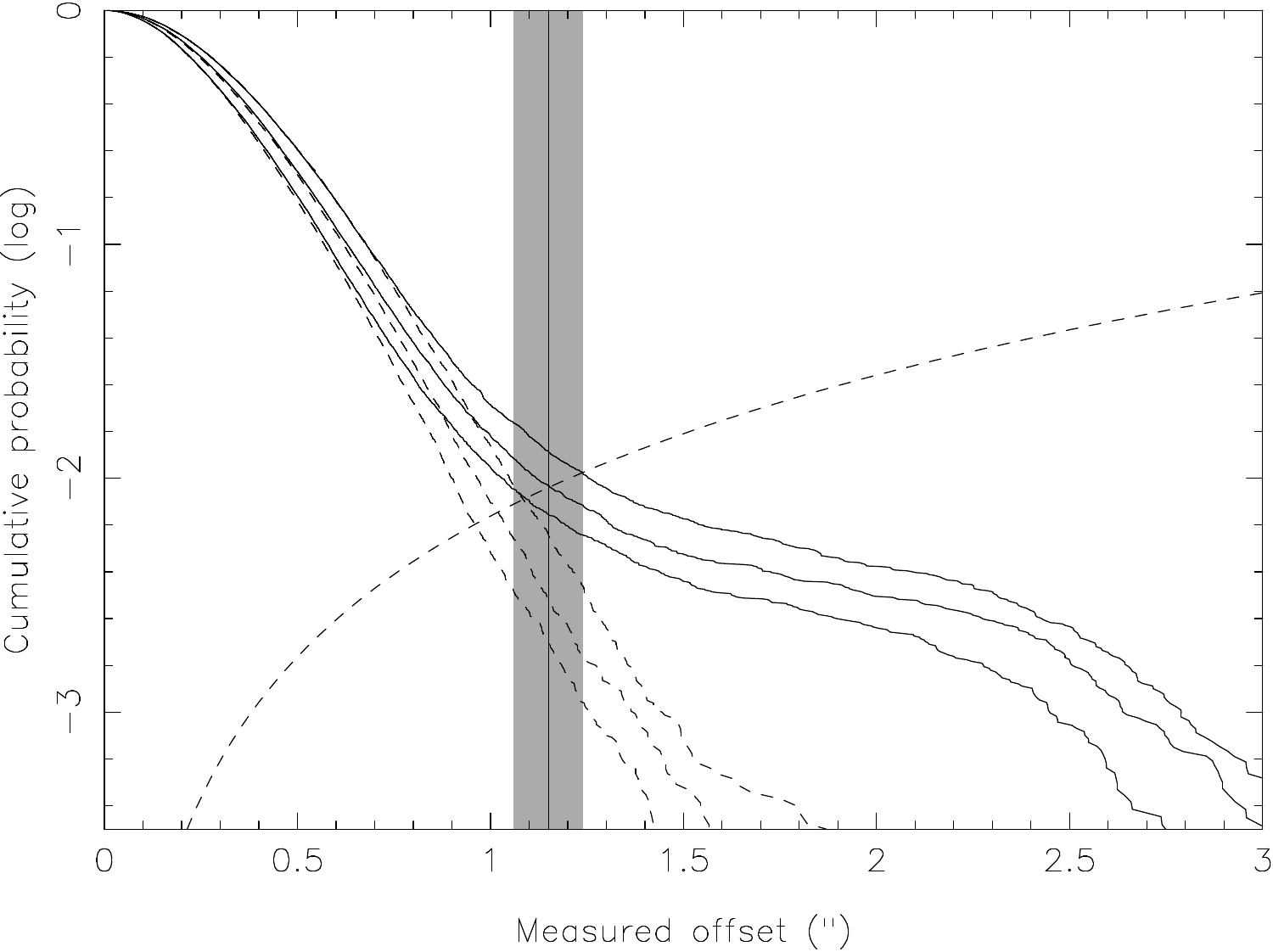}
\caption{Probabilities of finding the detected source more than a certain distance from the optical source, shown by the solid lines, which from top to bottom represent the assumed 1.2$''$, 1.0$''$ and 0.8$''$ PSF widths in the progenitor X-ray images. Dotted lines are the same but for artificial, pure-Poissonian backgrounds. The dashed line shows the trial-corrected chance alignment probability, and the solid vertical line the observed offset between the optical and X-ray source.}
\label{fig:positionerror}
\end{figure}

The resulting images were again analysed with \wav, and the positional offsets between the artificial sources and the reconstructed artificial sources were recorded. Fig.~\ref{fig:positionerror} shows the cumulative distribution of these offsets. The probability of finding a source a distance \dist or more away from the input source is low, close to 1\% for the 1.0$''$ PSF, and slightly higher and lower for the larger and smaller PSFs, respectively.

An interesting feature of the probability distributions shown in Fig.~\ref{fig:positionerror} are the multiple components: a component that drops off quickly on roughly the pixel size scale, and a much slower component. Presumably, this slower component represents cases in which the new source combines with an undetected source or overdensity in the nearby background, pushing it above the detection threshold, and being detected as a source offset from the input source. To test this hypothesis we ran a further Monte Carlo simulation, this time with purely Poissonian backgrounds of the same average value (within 10\% of 6.9 counts per 2$''$ aperture). The results are shown as the dotted lines in Fig.\ \ref{fig:positionerror}. Indeed, with the artificial backgrounds the slow component vanishes and the number of cases in which an offset of \dist\ or more is recorded drops, to about 0.3\% for a 1.0$''$ PSF.

In order to test whether the \wav\ algorithm could be responsible for some of the positional scatter in the Monte Carlo simulation, we repeated the analysis with a simple centroid algorithm, which we again applied only to simulated sources that were first detected with \wav\ in order to ensure that they are still representative of sources that would have been found in the original search by \citet{vossnature}. We measured centroids for the $N=\textrm{40,000}$ set of simulated images with a 1.0$''$ PSF. The amount of scatter in the centroid positions depends on the aperture chosen: for a 2$''$ aperture it is slightly smaller, while for 1$''$ and 3$''$ apertures it is slightly larger than the \wav\ scatter. An adaptive aperture, starting from 3$''$ and decreasing in 10\% steps to 1.5$''$ after positional convergence in each step, gives the same scatter as a 2$''$ aperture, suggesting that this is about the best one can do (in minimizing the positional scatter). For this `optimal' aperture, the X-ray source centroid is offset from the optical SN by 1.09$''$, slightly less than the \wav\ offset of \dist. The net result is that the offset probability is very similar for centroid and \wav\ positions, at approximately 1\%.

The important question now is how these numbers compare to the probabilities of a chance alignment of the optical SN with an X-ray source in the field. For this we counted the number of X-ray sources in the progenitor image found by \wav. We filtered out the region around NGC 1404 where the background is within 50\% of the value measured around 2007on, which should be representative of both the position where a SN would typically be discovered, and of the probability of detecting an X-ray source there with \wav. In an area measuring 7.1 square arcmin \wav\ finds 17 sources, giving a probability of 0.28\% for a chance alignment within \dist\ of the optical position of 2007on. Note that this figure somewhat conservatively includes sources that are detected with a lower significance level than the possible progenitor of 2007on. Counting only equally or more significant detections gives 14 sources, or a 0.23\% chance alignment probability.

Given that SN 2007on was the fourth trial in their search for progenitor detections in archival \emph{Chandra} data \citep{vossnature}, one has to conclude that the chance alignment probability for the detected source is $\sim$0.9\%, or $\sim$1.1\% for any detectable source.

\section{Discussion}
\label{discussionsection}

Based on just the astrometry, the chances of the X-ray source being related to the supernova are small but equal to the chances of a chance alignment. The fact that the X-ray source appears to have dimmed (or disappeared) in the \emph{Chandra} DDT observations suggests that at least some of the X-ray photons in the progenitor images may have come from the progenitor, although alternatively, the X-ray source might be intrinsically variable (and unrelated). In this section we try to collect further evidence that may point to a relation (or not) between the X-ray source and the supernova.

\subsection{Soft versus hard X-ray photons}

\citet{vossnature} mentioned the softness of the X-ray source close to 2007on as further suggestive evidence of a connection between the two. In the accreting models, the accretion of matter by a white dwarf at a high rate is expected to create a strong, soft X-ray source (e.g.\ \citealt{vandenheuvel}), which makes the hardness of the source a potentially useful property.

\begin{figure}
\includegraphics[width=\columnwidth]{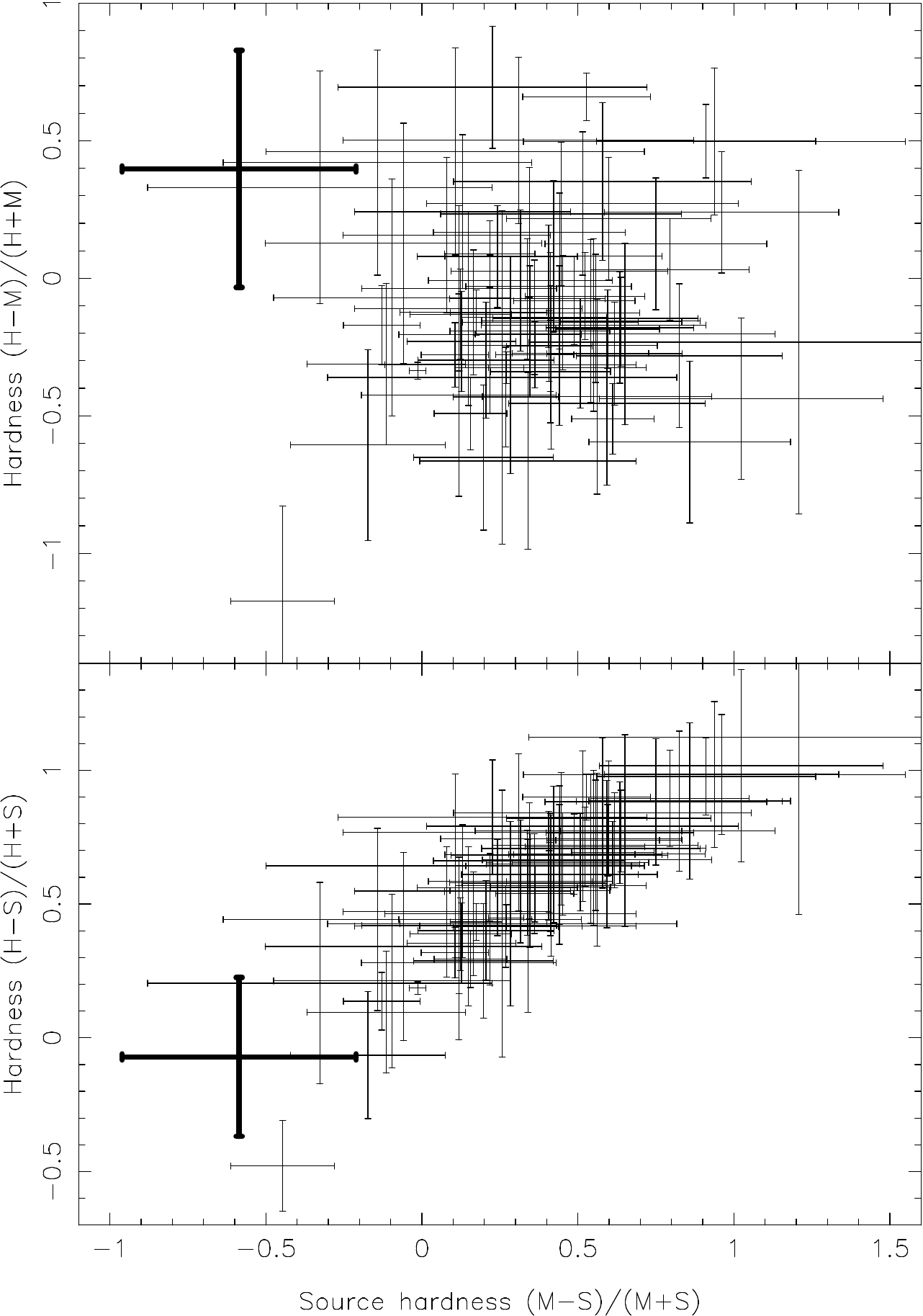}
\caption{Hardness ratios of X-ray sources in the field of 2007on, measured in photons corrected for the background flux levels. The S, M and H energy bands correspond to $0.3-1.0$\,keV, $1.0-2.0$\,keV and $2.0-8.0$\,keV, respectively. The source close to 2007on is the left-most data point in both panels, shown in bold.}
\label{fig:hardness}
\end{figure}

Fig.~\ref{fig:hardness} shows the hardness ratios of all X-ray sources in the combined pre-outburst image that are detected with at least 15 photons. Photons are binned into S(oft), M(edium) and H(ard) energy bands, corresponding to $0.3-1.0$\,keV, $1.0-2.0$\,keV and $2.0-8.0$\,keV photon energies, respectively, and colours are constructed following the method of \citet{prestwich}. The source close to 2007on is the softest source in terms of its soft-to-medium hardness ratio, while its medium-to-hard hardness ratio is quite hard and more comparable to the other sources in the field. Its spectrum is thus difficult to interpret, but a possible explanation could be that there are two components to the spectrum: a very soft one and a rather hard one. Although a few extra photons in the M band would have pushed the source in the direction of the more `mundane' region occupied by the majority of sources in the upper panel of Fig.\ \ref{fig:hardness}, the source would still have shown an overabundance of soft X-rays, as is clear from the lower panel in this figure.

\begin{figure*}
\includegraphics[width=\columnwidth]{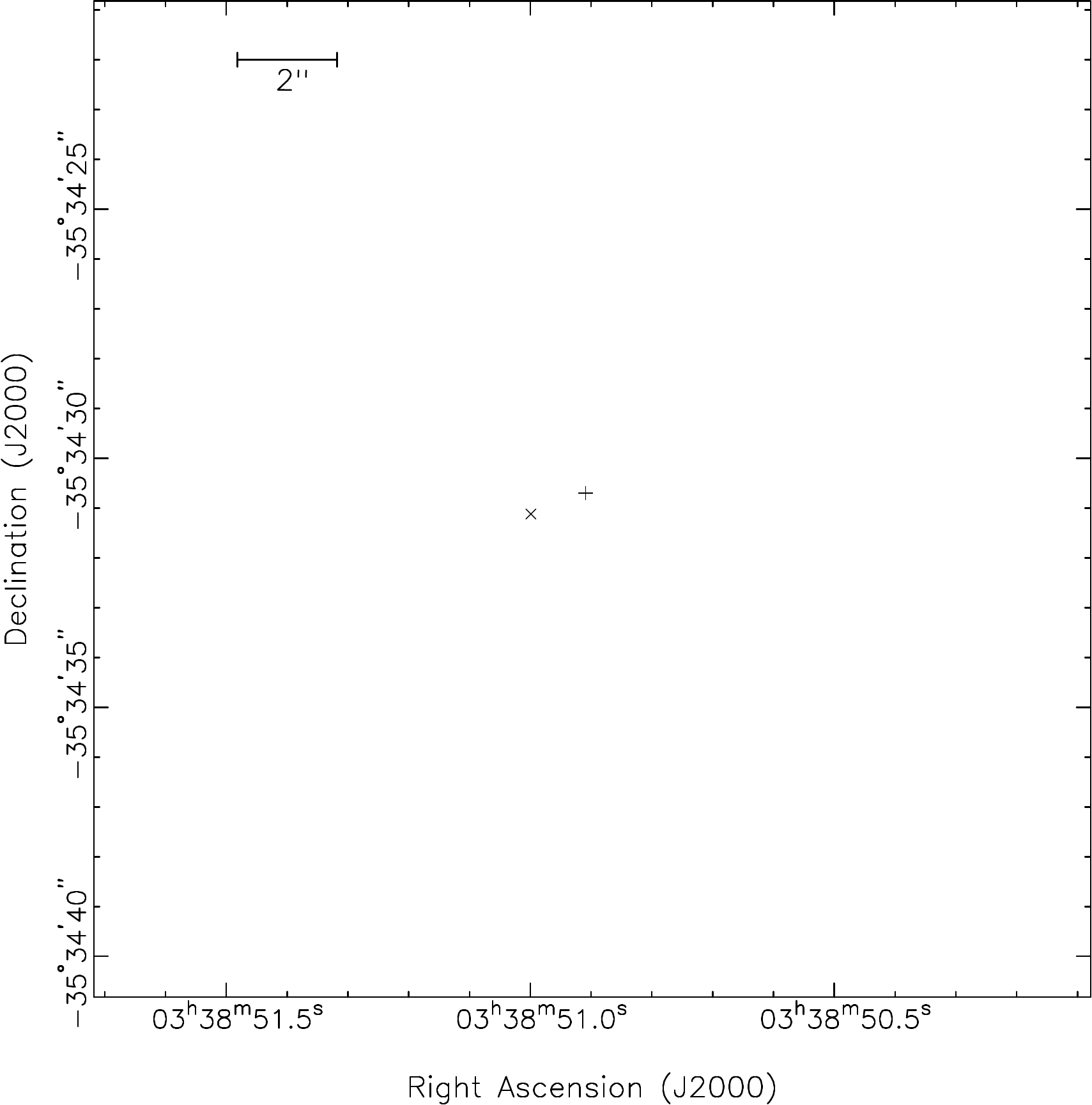}\hfill%
\includegraphics[width=\columnwidth]{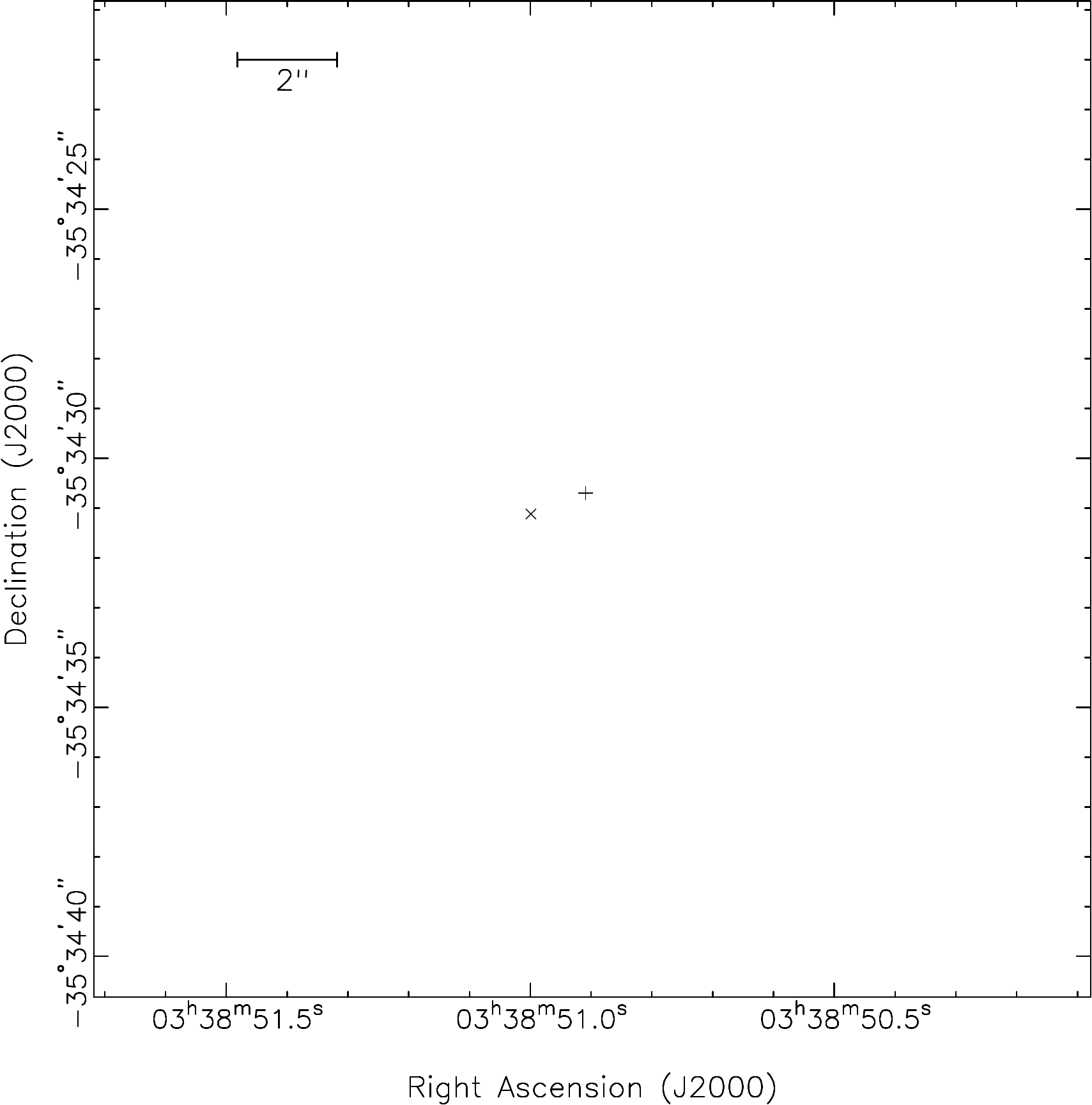}
\caption{Pre-outburst \emph{Chandra} image split into soft ($0.3-1.0$\,keV, left panel) and medium/hard ($1.0-8.0$\,keV, right panel) photons. The positions of the optical SN ($\times$) and the X-ray source as reconstructed by \wav\ for the combined image ($+$) are indicated.}
\label{fig:positions_energies}
\end{figure*}

To further investigate the possibility of a two-component source, we show in Fig.~\ref{fig:positions_energies} the pre-outburst X-ray image divided into soft ($0.3-1.0$\,keV) and medium/hard ($1.0-8.0$\,keV) photons, together with the position of the optical supernova. The obvious problem is that the photon statistics of these images become even worse. Nevertheless the medium/hard photons are seen to line up with the position of the reconstructed X-ray source, while the soft photons appear more scattered. Although the photons closest to the optical SN are soft, there is no general trend by which the soft photons line up better with the optical position. The figure does therefore not provide convincing evidence for a connection between the optical supernova and the soft X-rays in particular.

\subsection{\emph{Hubble Space Telescope} images}
\label{hstsection}

\begin{figure}
\includegraphics[width=\columnwidth]{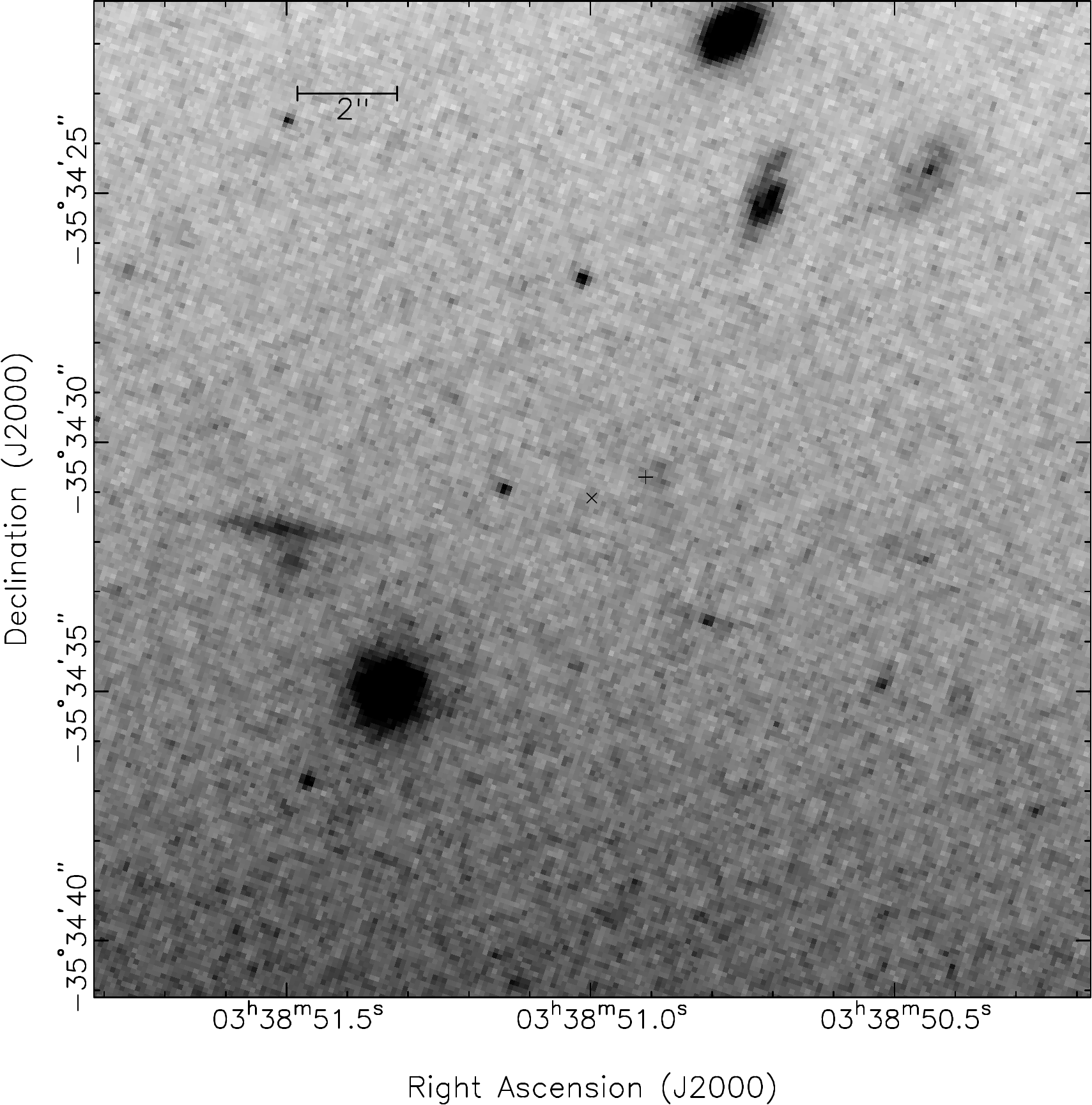}
\caption{Drizzled \emph{HST}/ACS image of the region around 2007on, totalling 1224\,s in the far-red F814W filter. A faint, extended object appears to be present near the X-ray source position from \wav\ ($+$), although it is close to the detection limit of $I\sim26.5$. Nothing is visible at the position of the optical supernova ($\times$).}
\label{fig:positions_hst}
\end{figure}

Given the indication that the progenitor X-ray source may (partly) have been an unrelated, nearby source, we reanalyse the archival \emph{HST} images of the region around 2007on, while performing the same accurate astrometry. Figure \ref{fig:positions_hst} shows a drizzled \citep{fruchter} 1224-s image taken August 6, 2006 with the Advanced Camera for Surveys (ACS) using the F814W far-red filter. There is evidence for a faint, extended source close to the position of the X-ray source, although it is close to the detection limit of $I\sim 26.5$. Nothing is visible at the optical position of the SN.

If we compare the observed X-ray flux $f_X\sim2\times 10^{-15}$\,erg\,s$^{-1}$\,cm$^{-2}$ in the $0.3-8.0$\,keV band (see \citealt{2007sr}, after a correction to the value given in \citealt{vossnature}) to the $I$-band flux $f_I$ of the object in the ACS image, and we compare this to results from the \emph{Chandra} Deep Field North \& South surveys \citep{alexander,giacconi}, we conclude that the observed flux ratio $f_X/f_I \gtrsim 10$ is at or beyond the upper ends of the distributions observed in those surveys. This makes it unlikely that all of the observed X-ray photons originate from this possible background source, although it is conceivable that a few of them do. Nevertheless we conclude that an origin within NGC\,1404 for the X-ray source appears to be more likely.

\section{Conclusion}
\label{conclusionsection}

We have presented accurate astrometry of the field of type-Ia supernova 2007on, and conclude that there appears to be an offset of \disterr\ between the optical supernova and the X-ray source close to it in archival \emph{Chandra} images. Based on just this information, and barring astrometric errors, the probability of the X-ray source being related to the supernova is small: about 1\%. The probability for a chance alignment with an unrelated X-ray source is, unfortunately, equally small, which makes it difficult to decide between the two scenarios.

Several additional pieces of information suggest that at least part of the X-ray source may have been related to the supernova, although none of them is decisive. First there is the indication that the X-ray source has dimmed (or disappeared) in the post-outburst \emph{Chandra} images taken about six weeks after optical maximum, at the 90--99\% confidence level depending on the choice of aperture size. Second, the source in the pre-outburst images shows an excess of soft X-ray photons relative to the other sources in the field, and SN Ia progenitors in the accreting models are expected to be prolific sources of soft X-rays. One could argue that the probability for a chance alignment between the optical SN and a \emph{soft} X-ray source is much lower than the quoted 1\% probability of an alignment with \emph{any} X-ray source, which would make a physical connection between the X-ray source and the optical SN \emph{less unlikely} than a chance alignment. A possible extended background object close to the X-ray source position (but away from the optical supernova), as seen in archival \emph{HST}/ACS data, could conceivably be responsible for some of the observed hard X-rays.

Probably the only way out of the impasse will be future \emph{Chandra} observations of 2007on, which can decide whether the X-ray source close to 2007on really has dimmed or disappeared. X-rays from the supernova are not expected to be detectable with \emph{Chandra}: \citet{immlerpaper} consider Compton scattering of gamma rays from radioactive decay which, although model-dependent, is expected to fall short of the detection limit by three orders of magnitude, while the X-ray luminosity from interaction of the SN ejecta with the local interstellar medium is expected to fall off as 1/time and should thus have decreased by an order of magnitude since our DDT observations. A future detection would thus imply that the source is unrelated. Although the possibility of the X-ray source having been an unrelated transient source can never be ruled out completely, a future non-detection would further lower the probability of the source having been a chance alignment, since only a few per cent of all X-ray sources observed in a given extragalactic snapshot are transients \citep{vossgilfanov}.

\section*{Acknowledgments}

GR is supported by NWO Rubicon grant 680.50.0610 to G.H.A. Roelofs. RV was supported by the DFG cluster of excellence `Origin and Structure of the Universe'. GN is supported by the Netherlands Organisation for Scientific Research. We thank the director of the Chandra X-ray Observatory for allocating Director's Discretionary Time, and the referee for suggesting to redo our analysis without the standard photon position randomization. This research has made use of data obtained from the Chandra Data Archive and
software provided by the Chandra X-ray Center in the application package \textsc{ciao}.
Based on observations made with the NASA/ESA Hubble Space Telescope, obtained
from the data archive at the Space Telescope Science Institute. STScI is
operated by the Association of Universities for Research in Astronomy, Inc.\
under NASA contract NAS 5-26555. Based on data obtained
from the ESO Science Archive Facility. Based on observations made with ESO
Telescopes at the La Silla Observatory under programme
080.A-0516.

{}

\end{document}